\documentstyle[12pt,epsf]{article}

\def\fun#1#2{\lower3.6pt\vbox{\baselineskip0pt\lineskip.9pt
        \ialign{$\mathsurround=0pt#1\hfill##\hfil$\crcr#2\crcr\sim\crcr}}}

\renewcommand\({\left(}
\renewcommand\){\right)}
\renewcommand\[{\left[}
\renewcommand\]{\right]}

\newcommand\eq[1]{Eq.~(\ref{#1})}
\newcommand\eqs[2]{Eqs.~(\ref{#1}) and (\ref{#2})}

\newcommand\ee{\end{equation}}
\newcommand\be{\begin{equation}}
\newcommand\eea{\end{eqnarray}}
\newcommand\bea{\begin{eqnarray}}

\newcommand\GeV{\,\mbox{GeV}}
\newcommand\MeV{\,\mbox{MeV}}

\newcommand\mpl{M_{\rm P}}

\newcommand\lsim{\mathrel{\rlap{\lower4pt\hbox{\hskip1pt$\sim$}}
    \raise1pt\hbox{$<$}}}
\newcommand\gsim{\mathrel{\rlap{\lower4pt\hbox{\hskip1pt$\sim$}}
    \raise1pt\hbox{$>$}}}

\def\dslash{\not{\hbox{\kern-2pt $\partial$}}}
\def\Dslash{\not{\hbox{\kern-4pt $D$}}}
\def\Oslash{\not{\hbox{\kern-4pt $O$}}}
\def\Qslash{\not{\hbox{\kern-4pt $Q$}}}
\def\pslash{\not{\hbox{\kern-2.3pt $p$}}}
\def\kslash{\not{\hbox{\kern-2.3pt $k$}}}
\def\qslash{\not{\hbox{\kern-2.3pt $q$}}}

 \newtoks\slashfraction
 \slashfraction={.13}
 \def\slash#1{\setbox0\hbox{$ #1 $}
 \setbox0\hbox to \the\slashfraction\wd0{\hss \box0}/\box0 }
 
\def\ee{\end{equation}}
\def\be{\begin{equation}}

\newcommand\sub[1]{_{\rm #1}}

\begin{document}

\begin{flushright}
LANCS-TH/9820
\\hep-ph/9809562\\
(September 1998)
\end{flushright}
\begin{center}
{\Large \bf Running-mass models of inflation, and their observational
constraints}

\vspace{.3in}
{\large\bf  Laura Covi and David H.~Lyth}

\vspace{.4 cm}
{\em Department of Physics,\\
Lancaster University,\\
Lancaster LA1 4YB.~~~U.~K.}

\vspace{.4cm}
{\tt E-mails: l.covi@lancaster.ac.uk\ d.lyth@lancaster.ac.uk \
}
\end{center}

\vspace{.6cm}
\begin{abstract}
If the inflaton sector is described by softly broken supersymmetry,
and the inflaton has unsuppressed couplings, the inflaton mass will
run strongly with scale. Four types of model are possible. 
The prediction for the spectral index involves two parameters,
while the COBE normalization involves a third, all of them calculable 
functions of the relevant masses and couplings. A crude estimate is made 
of the region of parameter space allowed by present observation.
\end{abstract}

\section{Introduction}

It is generally supposed that 
inflation sets the initial conditions for the subsequent
hot big bang. In particular the 
primordial density perturbation, that is thought to be the origin of 
structure in the Universe, is supposed to come from the vacuum 
fluctuation of the inflaton field. The spectrum
$\delta\sub H^2$
of the primordial density perturbation
involves only the potential $V(\phi)$
of the inflaton field $\phi$, evaluated while cosmological scales are 
leaving the horizon.
It is given by
\be
\delta_H^2(k) = \frac1{ 75\pi^2 \mpl^6}\frac{V^3}{V'^2} \,,
\label{delh}
\ee
where $k/a$ is the comoving wavenumber,
and 
the right hand side is evaluated at the epoch of horizon exit
$k=aH$.\footnote
{The Planck scale is $\mpl = (8\pi G)^{-1/2} =2.4 \times 10^{18}\GeV$.
The scale factor of the Universe is $a$, normalized to $a=1$ at the 
present epoch,
and $H\equiv \dot a/a$ defines the Hubble parameter, with a dot
indicating the time-derivative. }
The scale-dependence of the spectrum is conveniently specified by a 
spectral index $n=1+d\ln \delta_H^2/d\ln k$, for which the 
prediction is\footnote
{The precise slow-roll result is $\frac12(n-1) = \eta -\frac32
\mpl^2(V'/V)^2$, but the second term is negligible in most models
including the running mass ones.
The reader is referred to reference 
\cite{p97toni} for a comprehensive 
review of inflation models, including 
many details and references omitted in the 
present brief account.}
\be
\frac{n-1}2 = \eta\equiv \mpl^2 V''/V \,.
\label{nofeta}
\ee

On a scale $k\simeq 10 H_0$ ($H_0$ being the present 
value of $H$), the COBE observation of the cosmic 
microwave background anisotropy gives 
\be
\delta_H =1.91\times 10^{-5} \,.
\label{cobenorm}
\ee
A variety of observations indicate that $\delta_H(k)$ is approximately 
scale-independent over the cosmological range of scales
$H_0 \lsim k\lsim 10^4 H_0$.
The constraint assuming roughly constant
$n$ is \cite{abook}
\be
|n-1|\lsim 0.2 \,.
\label{ncon}
\ee

Observations of the cosmic microwave background anisotropy
and of galaxies, plus accurate determinations of the cosmological
parameters, will strongly discriminate between
models of inflation within the next decade or so.
In particular, the Planck satellite 
\cite{planck} will eventually measure $n$ with an 
accuracy $\Delta n\sim 0.01$.
Partly for this reason, the level of activity in
inflation model-building is quite high at present, and likely
to become higher over the next few years. In this paper we propose
a strategy for comparing with observation a whole class of models.
These are the models with a running inflaton mass 
\cite{ewanloop1,ewanloop2,p98laura}.

Let us begin by placing this class of models in perspective.
For many years, the standard paradigm
was the tree-level potential
\be
V(\phi) = V_0 + \frac12 m^2 \phi^2 + \frac14 \lambda \phi^4 + \cdots \,.
\label{vtree}
\ee
The dots represent non-renormalizable terms, and the constants
$V_0$, $m^2$ and $\lambda$ are supposed to have negligible
dependence on the renormalization scale (no running).
For the underlying field theory to be under control one needs
$\phi\lsim \mpl$, and we focus on this case. Then inflation requires
a very flat potential, with $V_0$ dominating. Observation 
requires \cite{p97toni}
\be
\mpl^2 m^2/V_0 \lsim 0.1 \,,
\label{mcon}
\ee
and 
\be
\lambda \lsim 10^{-8} \,.
\label{lamcon}
\ee
Because of the last constraint, $\lambda$ is usually supposed to be 
completely negligible during inflation, along with the non-renormalizable terms.
This leaves the mass term, giving the distinctive tree-level 
prediction\footnote 
{In writing the tree-level potential, we ignored odd terms
on the assumption that they are forbidden by some symmetry.
We also ignored terms with negative or fractional powers, that arise in
mutated hybrid inflation models, where the non-inflaton field 
has a $\phi$-dependent value. If any such term dominates, it gives
a spectral index fairly close to 1, with the mild scale-dependence
$n-1\propto 1/N$.}
that $n$ is scale-independent and possibly indistinguishable from 1.
We note for future reference that a generic supergravity theory
makes all scalar field masses, and in particular the inflaton mass,
of order
\be
|m| \sim V_0^{1/2}/\mpl \,,
\label{sugram}
\ee
in contradiction with the bound
\eq{mcon}.

Because the tree-level potential is so flat, the one-loop correction
may in fact be significant. In the regime where  $\phi$ is much bigger 
than all of the relevant masses, it
typically has the form
$\Delta V=(a_0+a_2\phi^2 + a_4\phi^4)
\ln(\phi/Q)$. Here $Q$ is 
the renormalization scale, which should be fixed at some
value within the relevant range of $\phi$ so that the loop correction
is small (and therefore believable).

In a non-supersymmetric theory, the quartic term dominates and
each loop gives a contribution 
\be
\Delta V = g^2\phi^4 \ln(\phi/Q) \,,
\ee
where $g\ll 1$ is a typical coupling (times a loop-suppression factor
like $(8\pi)^{-1/2}$).
This is roughly equivalent to a change
$\Delta\lambda \sim g^2$ in the tree-level
potential, so the constraint 
\eq{lamcon} requires very suppressed couplings ($g^2$ many orders of 
magnitude below 1). In particular
the inflaton is presumably a gauge singlet, 
since gauge couplings as opposed to Yukawa couplings 
are not supposed to be suppressed.

With the favoured paradigm of supersymmetry, things are quite different.
During inflation, the sector of the theory occupied by the
inflaton field is usually described by global supersymmetry as opposed 
to supergravity.\footnote
{By `sector' we mean the set of fields that couple to the inflaton with 
more than gravitational strength.}
The quartic term then vanishes by virtue of non-renormalization theorems.
If supersymmetry is broken only spontaneously, and the relevant
masses-squared have zero supertrace, the quadratic term vanishes as well
leaving a contribution of the form
\be
\Delta V = g^2 V_0 \ln(\phi/Q) \,.
\ee
Again, $g$ is a typical coupling. 
On the assumption that the slope of the tree-level potential is 
negligible, this paradigm has been widely studied, particularly
in the manifestation known as `$D$-term inflation'. It makes the
distinctive prediction $n\simeq 0.97$ which can eventually be tested.
Unfortunately, it also (through \eq{cobenorm}) requires
$\phi\sub{COBE}\gsim\mpl$ if $g$ is unsuppressed, which is typically the 
case in this type of model. (A subscript COBE will indicate the epoch 
when the scale explored by COBE leaves the horizon.)
As a result the non-renormalizable terms
of the tree-level potential will generically spoil inflation.

We are here concerned with the opposite possibility, that supersymmetry
during inflation is broken explicitly as opposed to spontaneously.
As usual the breaking is supposed to be soft, and then the dominant
loop correction is quadratic,
\be
\Delta V = g^2 \tilde m^2 \phi^2 \ln(\phi/Q) \,.
\label{dV-soft}
\ee
Here $\tilde m$ 
is a typical soft mass appearing in the loop,
and $g$ is still a typical coupling. 
Again, the loop correction is significant if $g$ is 
unsuppressed, assuming that $|\tilde m|$ has the typical value
$V_0^{1/2}/\mpl$.
We assume that these conditions are satisfied in what follows.

With this paradigm, $\phi$ typically varies by many orders of magnitude 
during the era of interest (starting when
COBE scales leave the horizon, and ending when slow-roll inflation
ends). 
As a result, no single choice of $Q$ will make the loop correction valid 
during the entire era. To handle this situation, one can drop the 
loop correction in favour of
the renormalization-group-improved 
tree-level potential, in which $m^2(Q)$ is evaluated at 
$Q=\phi$;
\be
V = V_0 + \frac12m^2(\phi) \phi^2 + \cdots \,.
\label{vrun}
\ee
In the 
approximation that $m^2$ is linear in $\ln\phi$, one recovers the prescription
`unimproved 
tree-level plus loop correction' by fixing the renormalization scale.

Following Stewart \cite{ewanloop1,ewanloop2}, our central strategy
is to assume the linear approximation while cosmological scales
are leaving the horizon, but not afterwards. The 
predictions then involve only three parameters, which we call
$c$, $\sigma$ and $\tau$. The first two give the prediction for the
spectral index, while the COBE prediction involves all three.

The plan of the paper is as follows. In Section \ref{s2} we 
evaluate the predictions, and the four possible types of model
to which they apply. In Section \ref{s3} we make a crude estimate of
the observational constraints on the three parameters,
for each of the four possible models.
In Section \ref{s4} we see how the present discussion applies to
the simplest possible model \cite{ewanloop2,p98laura}, 
in which only a single gauge coupling is significant.
In Section \ref{s5} we summarize the results,
and point to future directions for comparison with observation.

\section{The running-mass models and their predictions}

\label{s2}

The generic estimate \eq{sugram} applies 
to all of the scalar masses, when the renormalization scale is
$\mpl$. As Stewart pointed out, it 
can be avoided for the inflaton provided that 
$m^2(\phi)$ decreases in magnitude, as $\phi$ decreases from $\mpl$. 
We assume that this happens, and focus on the case
\cite{ewanloop1,ewanloop2} where $m^2(\phi)$ 
changes sign at some point.
(The opposite case will be mentioned briefly.)
Because the couplings are small compared with unity, 
$V'$ then vanishes at some relatively nearby point, which we 
denote by $\phi_*$. 

\subsection{The linear approximation}

It is useful to write \eq{vrun} in the form
\be
V(\phi) = V_0 \( 1 - \frac12 \mpl^{-2} \mu^2(\phi) \phi^2 \) \,,
\ee
where
\be
\mu^2(\phi) \equiv -\mpl^2 m^2(\phi)/V_0 \,.
\ee
We are supposing that $V_0$ dominates, since this is necessary 
for inflation in the regime $\phi\lsim \mpl$ where the field
theory is under control.
Then
\bea
\mpl\frac{V'}{V_0} 
&=& - \phi \left[\mu^2 + {1\over 2} {d\mu^2\over dt} 
\right] 
\label{V-prime}\\
\eta\equiv 
\mpl^2\frac{V''}{V_0} &=& - 
\[ \mu^2 + {3\over 2} {d\mu^2\over dt} +
{1\over 2} {d^2\mu^2\over dt^2} \]
\,,
\label{eps-eta}
\eea
where $t\equiv\ln(\phi/\mpl)$.

We 
assume that while observable scales are leaving the 
horizon one can make a linear expansion in $\ln\phi$,
\be
\mu^2 \simeq \mu_*^2 + c\ln(\phi/\phi_*) \,,
\label{linear}
\ee
where $|c|\ll 1$ is related to the couplings involved.
This gives
\bea
\mpl \frac{V'}{V_0} &=& c\phi\ln (\phi_*/\phi) 
\label{vp} \\
\eta \equiv \mpl^2 \frac{V''}{V_0} 
&=&
c \[ \ln (\phi_*/\phi) -1 \] \,.
\label{vpp}
\eea
Note that $\mu_*^2=-\frac12c$, and that $\mu^2=0$ 
at $\ln(\phi_*/\phi) = -\frac12$ while $V''=0$ at 
$\ln(\phi_*/\phi) = 1$.

The number $N(\phi)$ of $e$-folds to the end of slow-roll inflation
is given by
\be
N(\phi) = \mpl^{-2}\int_{\phi\sub{end}}^\phi \frac{V}{V'} d\phi
\label{Nfull}
\,.
\ee

Using the linear approximation near $\phi_*$, this gives
\be
N(\phi) = -\frac1c \ln \( \frac c \sigma \ln \frac{\phi_*}{\phi} \) \,,
\label{N}
\ee
or
\be
(\sigma/c) e^{-cN} = \ln(\phi_*/\phi) \,.
\label{ecN}
\ee
Knowing the functional form of $m^2(\phi)$, and the value
of $\phi\sub{end}$, the constant $\sigma$ can be evaluated
by taking the limit $\phi\to\phi_*$ in
the full expression \eq{Nfull}. 
We shall see in the next section that in most cases 
one expects
\be
|c|\lsim |\sigma| \lsim 1 \,.
\label{sigmaexpect}
\ee

Using \eq{nofeta}, the
spectral index is given in terms of $c$ and $\sigma$
by 
\be
\frac{n(k)-1}{2} = \sigma e^{-cN} -c \,.
\ee
As usual, the right hand side is evaluated at the epoch of horizon exit
$k=aH$. Since $H$ is slowly varying, $d\ln k \simeq - dN$, where
$dN \equiv -d\ln a$ is the change in the number $N(\phi)$ of $e$-folds 
to the end of slow-roll inflation.

This prediction is relevant while cosmological scales are leaving
the horizon.
As we noted earlier, cosmological scales span roughly the four decades of 
$H_0\lsim k \lsim 10^4H_0$. Taking round 
figures, this corresponds to $\Delta N=10$, with the 
COBE measurement probing more or less the top of the range.
Cosmological scales therefore leave the horizon 
while 
\be
N\sub{COBE}- 10 \lsim N \lsim N\sub{COBE} 
\ee
To determine $N\sub{COBE}$ one needs to know what happens
after slow-roll inflation ends, but an upper bound is obtained by assuming
instant reheating which lasts until the present matter-dominated 
era. This gives\footnote
{This expression takes $k=H_0$ to be the scale probed by COBE. The more
correct $k\simeq 10H_0$ would reduce the right hand side by 
$\ln 10 \simeq2$.}
\be
N\sub{COBE} < 48 + \ln (V_0^{1/4}/10^{10}\GeV) \,.
\label{bound-N-V0}
\ee

From \eq{delh}, the 
COBE measurement gives 
\be
\frac{V_0^{1/2}}{\mpl^2} = 5.3\times 10^{-4} \mpl \frac{|V'|}{V_0} \,,
\ee
In our case it is convenient to define a constant $\tau$
by
\be
\ln(\mpl/\phi_*) \equiv \tau/|c|\,.
\ee

Assuming that $|m^2|$ has the typical value $V_0/\mpl^2$
at the Planck scale,
the linear approximation \eq{linear} 
applied at that scale would give $\tau\simeq 1$.
Will the linear approximation apply at that scale?
If {\em all} relevant masses at the Planck scale are of order 
$V_0/\mpl^2$, one expects on dimensional grounds that the linear 
approximation will be valid in the regime $|c\ln(\phi/\phi^*)|\ll
1$. Then the approximation will be just beginning to fail at the Planck 
scale. At least in this case, one expects $\tau$ to be very roughly of 
order 1.

Using the definition of $\tau$, 
\eqs{vp}{ecN} give
\be
\frac{V_0^{1/2}}{\mpl^2} = e^{-\tau/|c|}
\exp \(-\frac\sigma c e^{-cN\sub{COBE}} \) |\sigma | e^{-cN\sub{COBE}} 
\times 5.3\times 10^{-4} \,.
\label{thiscobenorm}
\ee

In these models, the spectral index may be strongly scale-dependent.
In fact, using $d\ln k=-dN$ one finds
\be
\frac{dn}{d\ln k} = 2 c\sigma e^{-cN} \,.
\ee
The observational constraint $|n-1|<0.2$ applies only if
$n$ has negligible variation,
$\Delta n\ll 0.4$.
Taking cosmological scales to span $\Delta \ln k\sim 10$,
and considering the centre of the range, the predicted variation is in 
fact negligible only if
\be
|c\sigma|e^{-c(N\sub{COBE}-5)}< 0.02 \,.
\label{negvar}
\ee

In a large region of parameter space, \eq{negvar} is violated
and $n$ has significant variation. In that 
situation, a detailed comparison with cosmological observations
is probably 
best done by directly considering the scale-dependence of
$\delta_H(k)$. Integrating $\frac12(n-1)
=d\ln \delta_H/d\ln k$, it is given by
\be
\ln \[\frac{\delta_H(k)}{\delta_H(k\sub{COBE})} \]
=+\frac\sigma c \( e^{-cN} - e^{-cN\sub{COBE} } \)
+c\( N- N\sub{COBE} \)
\,.
\label{delhvar}
\ee

The Planck satellite will explore a range
$\Delta \ln k\simeq 6$ of scales, and will measure $n$ with an accuracy
of around $.01$. This means that a scale-dependence 
$|dn/d\ln k|\gsim 2\times 10^{-3}$ should be observable by Planck.

\subsection{The four models}

Four types of inflation model are possible, corresponding to 
whether $\phi_*$ is a maximum 
or a minimum, and whether
$\phi$ during inflation is smaller or bigger than $\phi_*$.

\begin{figure}
\centering
\leavevmode\epsfysize=5.3cm \epsfbox{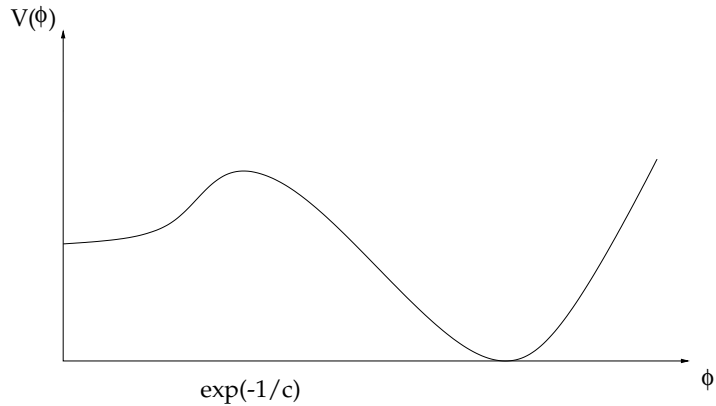}\\
\caption[loopc]{\label{f:loopc}
A possible form for the renormalization-group-improved inflaton
potential. In units of $\mpl$, the maximum is located roughly
at $\phi\sim e^{-1/c}$.
In the case illustrated, the minimum corresponds to the vacuum, where
$V$ vanishes.
This and the following figure are taken from 
\cite{p97toni}.}
\end{figure}

\begin{figure}
\centering
\leavevmode\epsfysize=5.3cm \epsfbox{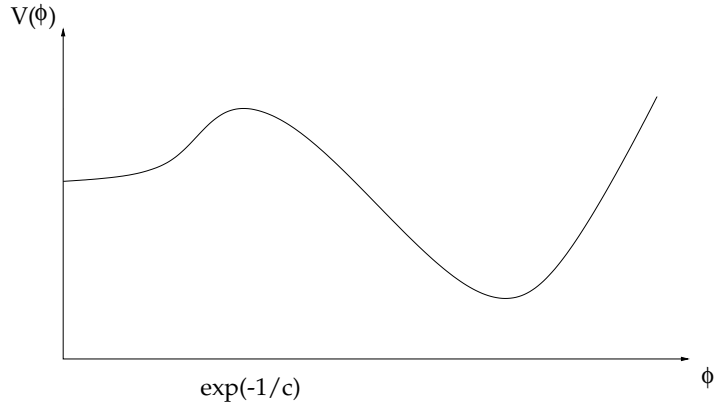}\\
\caption[loopd]{\label{f:loopd}
Alternatively, the true vacuum may lie in another field direction,
`out of the paper'.}
\end{figure}

If $\phi_*$ is a maximum, one expects the potential to have the form 
shown in Figures \ref{f:loopc} and \ref{f:loopd}. There is a minimum 
at $\phi=0$, and the non-renormalizable terms will ensure that there is
a minimum also at some value $\phi\sub{min}>\phi_*$.
The latter will generally be lower than the one at the origin, and we 
assume that this is the case. 
This lowest minimum represents the true vacuum if $V$ vanishes
there as in Figure \ref{f:loopc}. If instead  $V$ is positive
as in Figure \ref{f:loopd}, the vacuum lies in some other field
direction, `out of the paper'.

In the case that $\phi_*$ is a minimum, one expects that 
the origin will be a maximum, 
and that $V$ will increase monotonically to the right of the minimum.
The unique minimum represented by $\phi_*=0$ is the vacuum if $V$
vanishes there, otherwise the vacuum lies in some other field direction.

\subsubsection{Model (i); $\phi_*$ a maximum with $\phi<\phi_*$}

This model \cite{ewanloop2,p98laura} corresponds to 
$m^2(\mpl)<0$, 
$c>0$ and $\sigma>0$, with $\phi$ decreasing during inflation.
The spectral index increases as the scale $k^{-1}$ decreases,
and can be either bigger or less than 1.

For inflation to end, the form \eq{vrun} of $V(\phi)$ must
be modified when $\phi$ falls below some critical value
$\phi\sub c$, presumably through a hybrid inflation mechanism.
On the other hand, if the inflaton mass continues to run 
until $m^2\simeq V_0/\mpl^2$,
{\em slow-roll} inflation will end then.
Let us suppose first that this is the case, and define 
$\phi\sub{fast}$ by
\be
m^2(\phi\sub{fast}) = V_0/\mpl^2 \,.
\ee
This is equivalent to defining $\eta(\phi\sub{fast})=1$,
up to corrections of order $c$ which presumably should not be included 
in a one-loop calculation.
The end of
slow-roll inflation corresponds to $\phi\sub{end}=\phi\sub{fast}$,
and the linear approximation \eq{linear}
gives the rough estimate
$|\ln(\phi\sub{end}/\phi_*)|\sim 1/c$, making $\sigma\sim 1$.

Now consider the case where inflation ends at some value
$\phi\sub c$, with $|m^2(\phi\sub c)|<  V_0/\mpl^2$. If 
the mass is still running
at that point, the linear estimate \eq{N} gives
\linebreak
$\sigma \sim c\ln(\phi_*/\phi\sub c) < 1$. 
Values $\sigma\ll c$ can be achieved
only with $\phi\sub c$ very close to $\phi_*$ which would 
represent fine-tuning.
Therefore we expect in this case
$c\lsim \sigma\lsim 1$.

If the mass stops running before 
$\phi\sub c$ is reached, at some point $\phi\sub{low}$,
then $m^2$ has a constant value $m^2\sub{low}=m^2(\phi\sub{low})$ 
in the regime
$\phi\sub c < \phi <\phi\sub{low}$.
In this regime, some
number $\Delta N$ of $e$-folds of slow-roll inflation occur.
We are assuming that cosmological scales leave the horizon while the 
mass is still running, which requires
\bea
\Delta N &< & N\sub{COBE} - 10 \\
& <& 38 + \ln (V_0^{1/4}/10^{10}\GeV) \,.
\label{delN}
\eea
Retaining the estimate of the previous paragraph for
the $e$-folds of inflation before the mass stops running, the 
constant $\sigma$ to be used in \eq{ecN} will be in the range
\be
c\lsim \sigma \lsim e^{c\Delta N} \,.
\label{bigsigma}
\ee
According to the estimates of the next section, $ e^{c\Delta N}$ 
will not be more than one or two orders of magnitude above
unity. 

\subsubsection{Model (ii); $\phi_*$ a maximum with $\phi>\phi_*$}

Like the previous model, this one corresponds to $m^2(\mpl)<0$ and
$c>0$, but now $\sigma<0$ and $\phi$ increases during inflation.
The spectral index is less than 1, and decreases as the scale decreases.

In contrast with the previous case, inflation
can end without any need for  a hybrid inflation mechanism, or a change 
in the form of the potential \eq{vrun}, if the minimum at $\phi >\phi_*$
is the true vacuum. If the form \eq{vrun} holds until 
$\phi$ reaches the value $\phi\sub{fast}$ defined by
$\eta(\phi\sub{fast})=-1$, slow roll inflation will end there. 
To leading order in $c$ this corresponds
to\footnote
{This estimate of $\phi\sub{fast}$ assumes that quartic and 
higher terms in the tree-level potential
\eq{vtree} are negligible at $\phi\sub{fast}$. Assuming that only 
one such term is significant, one easily checks that
the estimate is roughly correct, unless the dimension 
of the term is not extremely large. We do not consider that case,
or the case where more than one term is significant.}
\be
m^2(\phi\sub{fast}) = -V_0/\mpl^2 \,.
\ee
Setting $\phi\sub{end}=\phi\sub{fast}$, and 
using the crude linear 
approximation
one finds $\phi\sub{end}\sim e^{1/|c|}\phi_*\sim \mpl$, and $\sigma\sim 
-1$.

On the other hand, slow-roll inflation might 
end at some 
point earlier $\phi\sub c$. In the true-vacuum case illustrated in
Figure \ref{f:loopc}, this may happen 
through a steepening in the form of $V(\phi)$. Otherwise it may happen 
through an inverted hybrid inflation mechanism. In both cases,
we expect $c\lsim |\sigma| \lsim 1$. 

In contrast with the previous model, this one also makes sense 
if $m^2$ stops running (as $\phi$ decreases)
before it changes sign; in other words,
if it stops running at $\phi\sub{low}$ with $m^2(\phi\sub{low})
<0$, but very small. In this case the maximum of the potential
is at the origin and $\eta$ is small and constant up to $\phi =0$.
The above treatment remains valid if $m^2$ has started to run before 
cosmological scales leave the horizon (remember that in this model,
$\phi$ increases during inflation).
Otherwise, one has a different model that we shall not consider.

\subsubsection{Model (iii); $\phi_*$ a minimum with $\phi<\phi_*$}

This corresponds to $m^2(\mpl)>0$, $c<0$ and $\sigma<0$,
and $\phi$ increases during inflation. The spectral index can
be either above or below 1, and it increases as the scale decreases.

Now $|m^2|$ decreases during 
inflation, and slow-roll inflation
ends only when
the potential \eq{vrun} 
ceases to hold at some value $\phi\sub{end}= \phi\sub c$.
In a single-field model, corresponding to $V$ vanishing at
 the minimum,
this 
can occur through a steepening of the form of the tree-level
potential, as higher powers of $\phi$ become important. Alternatively,
if $V$ is positive at the minimum
it can occur through a hybrid inflation mechanism (inverted hybrid
inflation).

To estimate $\sigma$ in this case, suppose first that 
(as $\phi$ decreases) the mass continues to run until 
$m^2= -V_0/\mpl^2$, and denote the point where this happens
by $\phi\sub{fast}$. Slow roll inflation can then 
only occur in the regime
$\phi\gsim \phi\sub{fast}$. 
It follows that 
\be
\phi\sub {end} \gsim \phi\sub{fast} \,,
\label{phiend}
\ee
and the 
linear approximation $\phi\sub{fast}\sim e^{-1/|c|}\phi_*$
then gives $|\sigma|\lsim 1$. As before $|\sigma|\gsim |c|$
is required to avoid the fine-tuning 
$\ln(\phi_*/\phi\sub c)\ll 1$.\footnote
{Stewart \cite{ewanloop1} took the view that models (iii) and (iv)
require a fine-tuning of $\phi\sub c$ over the whole range of parameter 
space. As with all views on fine-tuning, this is a matter of
taste.}

If the mass stops running at some point $\phi\sub{low}$,
with $|m^2(\phi\sub{low})|\ll 1$,
 inflation can begin at arbitrarily small field values.
If cosmological scales
start to leave the horizon only after the mass has started to run,
\eq{phiend} still applies and the estimate for
$\sigma$ is unchanged. We do not consider the opposite case.

\subsubsection*{Model (iv); $\phi_*$ a minimum with $\phi>\phi_*$}

Like the previous case this one corresponds to $m^2(\mpl)>0$
and $c<0$, but now $\sigma>0$ and $\phi$ decreases during inflation.
The spectral index is bigger than 1, and it decreases as the scale 
decreases.

Everything is the same as in the previous case, except that 
a hybrid inflation mechanism will definitely be needed to end
inflation, since higher-order terms in $\phi$ can hardly become
more important as $\phi$ decreases.
We again expect $|c|\lsim \sigma
\lsim 1$, with the lower limit needed to avoid the
fine-tuning $\ln(\phi\sub c/\phi_*)\ll 1$.
As a result we expect $|c|\lsim \sigma \lsim 1$.

Like Model (iii), this one can still make sense if the mass stops 
running before $\phi_*$ is reached. 
The above treatment applies if cosmological scales leave the horizon
while the mass is still running. We do not consider the opposite case.

\subsection{When will the mass stop running?}

The expression for the one-loop correction given in eq. (\ref{dV-soft})
is valid when $\phi$ is larger than any other mass scale and only 
in this case is the choice $Q=\phi$ reasonable. In the other case $Q$ 
will be given by some other relevant scale and the mass would no more 
depend on the value of $\phi$.
We can therefore assume that running stops 
at the value $\phi\sub{low}$, defined by
$\phi\sub{low}=|\tilde m(\phi\sub{low})|$, where $\tilde m$ is the 
biggest relevant mass.

At the Planck scale, the estimate \eq{sugram} applies to all 
masses.\footnote
{At least it applies to the scalar masses. Gaugino masses
can be smaller, depending on the gauge kinetic function.}
As the renormalization scale decreases, some will increase in 
magnitude but as a rough estimate we can take $|\tilde m(\phi)|$
to have the roughly scale-independent value
$V_0^{1/2}/\mpl$. According to the COBE normalization
\eq{roughcobe} given below, this gives
\bea
|\tilde m| & \sim &
(10^{-4}{\rm \ to \ }10^{-5} ) e^{-\tau/|c|} \mpl\\
&\equiv& (10^{-4}{\rm \ to \ }10^{-5} ) \phi_* \,.
\eea

According to this crude estimate, the running continues until well below
$\phi_*$, where $m^2\simeq 0$. In the
specific model \cite{p98laura} that we consider later,
it continues all the way down to $\phi\sub{fast}$,
where $|m^2|=V_0^{1/2}/\mpl$.

\section{Observational constraints}

\label{s3}

\begin{figure}
\centering
\leavevmode\epsfysize=6.5cm \epsfbox{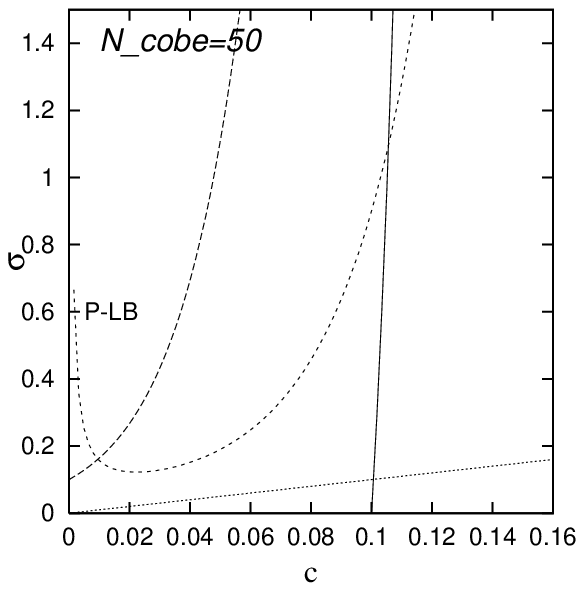}
\epsfysize=6.5cm \epsfbox{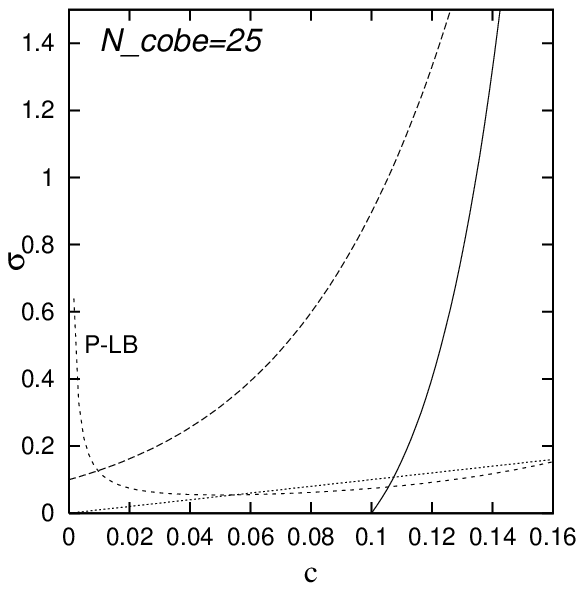}\\
\caption[modelone]{\label{f:modelone}
This and the following three figures show a crude observational 
constraint on the parameter space, obtained by requiring
$|n-1|<0.2$ at both $N=N\sub{COBE}$ and $N=N\sub{COBE}-10$.
These values of $N$ correspond respectively to the biggest and smallest
scales on which $n$ can be observed.
The first panel takes $N\sub{COBE}=50$, while the second takes
$N\sub{COBE}=25$. 
In all cases, a full line corresponds to $n=0.8$
while a long--dashed line corresponds to $n=1.2$.
The dotted straight line $ |\sigma | = |c|$ is 
also shown in all cases; since the theoretically expected 
parameters satisfy $ |\sigma | \geq |c|$, the allowed
parameter space is bounded by these three lines.
If $|\sigma | $ is bigger than the value indicated by
the short--dashed line (labelled P-LB), the Planck satellite 
will measure the scale dependence of $n$.}
\end{figure}

\begin{figure}
\centering
\leavevmode\epsfysize=6.5cm \epsfbox{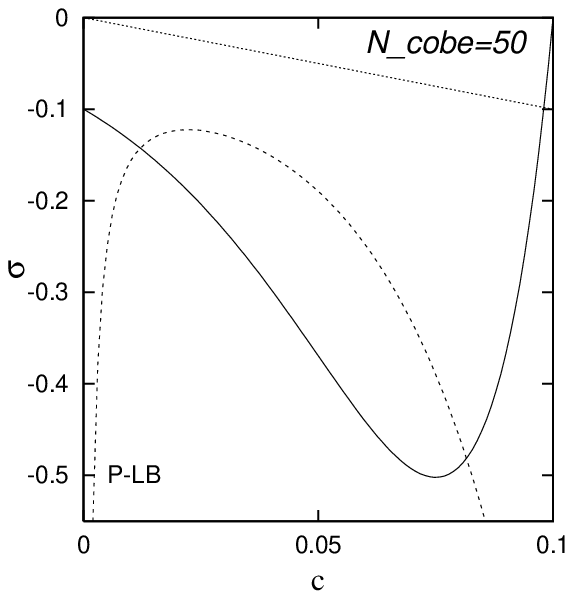}
\epsfysize=6.5cm \epsfbox{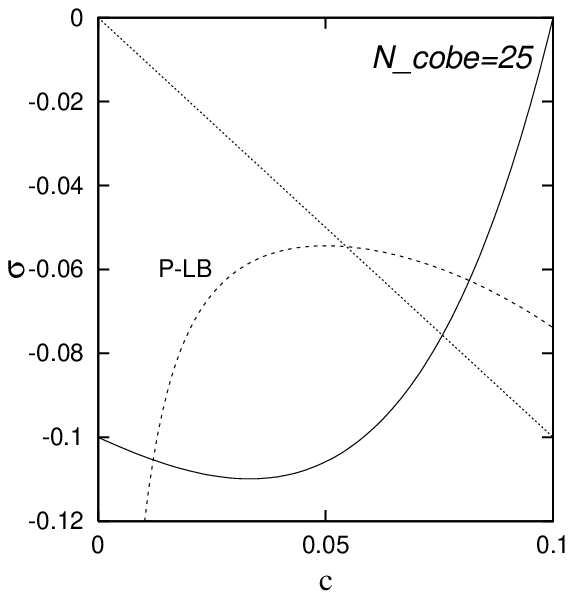}\\
\caption[modeltwo]{\label{f:modeltwo}
Model (ii), in which 
$n$ is less than 1, and  decreases as the scale decreases.
The  full line corresponds
to $n=0.8$ evaluated at $N=N\sub{COBE}-10$. The allowed region lies 
above this line and below the dotted one. 
For parameters lying below the short--dashed line (P--LB) 
the spectral index scale dependence will be detected by 
the Planck satellite.}
\end{figure}

\begin{figure}
\centering
\leavevmode\epsfysize=6.5cm \epsfbox{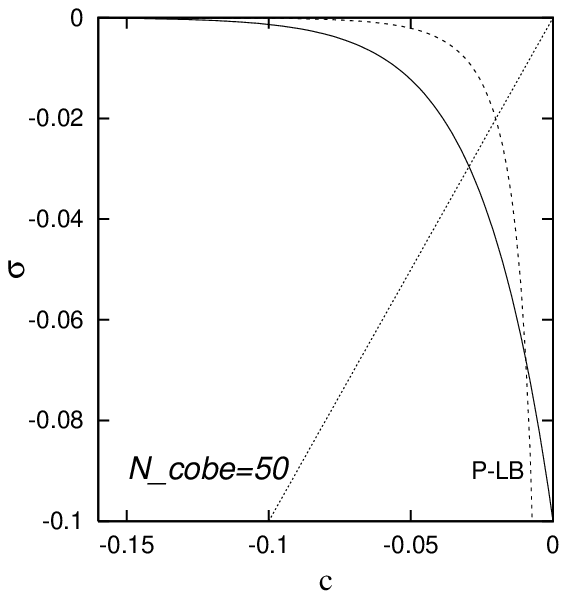}
\epsfysize=6.5cm \epsfbox{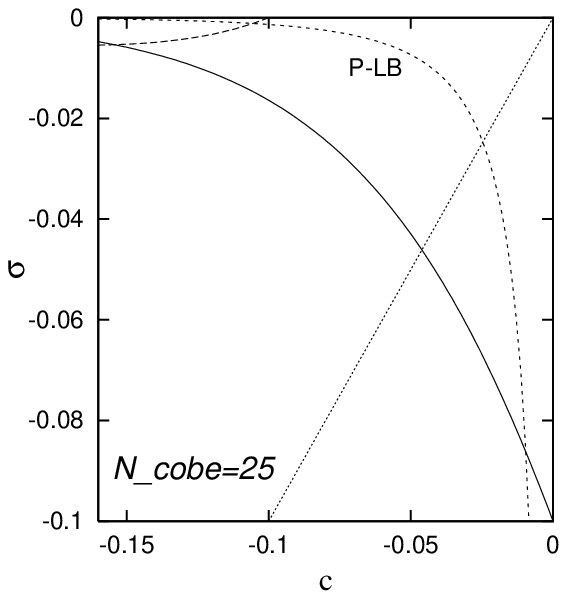}\\
\caption[modelthree]{\label{f:modelthree}
Model (iii), in which 
$n-1$ increases as the scale decreases.
The full line corresponds to $n=0.8$ evaluated
at $N=N\sub{COBE}$. 
The long--dashed line (which is invisible in the first 
panel) corresponds to
$n=1.2$ evaluated at $N\sub{COBE}-10$.
The allowed region lies between the two lines and the axes
and below the dotted line. 
For parameters lying below the short--dashed line (P--LB) 
the spectral index scale dependence will be detected by 
the Planck satellite.}
\end{figure}

\begin{figure}
\centering
\leavevmode\epsfysize=6.5cm \epsfbox{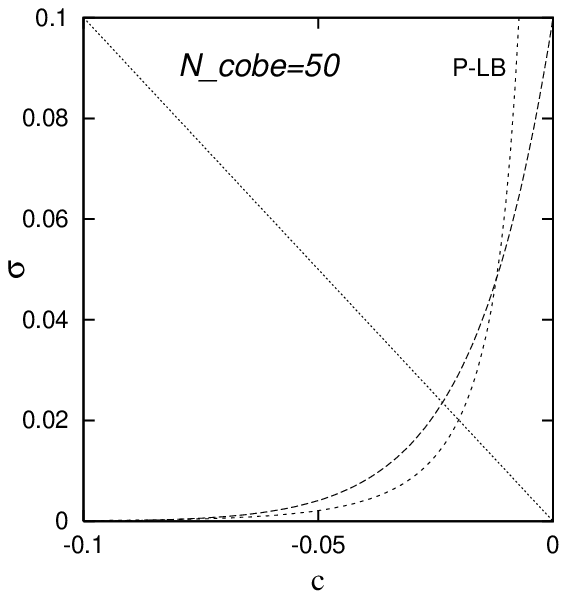}
\epsfysize=6.5cm \epsfbox{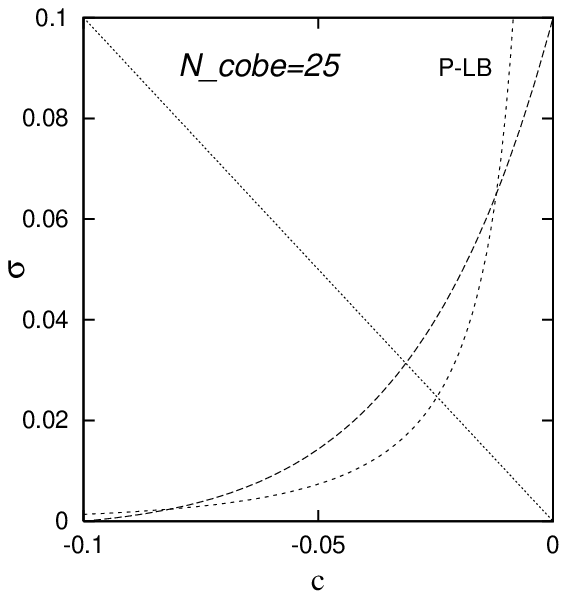}\\
\caption[modelfour]{\label{f:modelfour}
Model (iv), in which $n-1$ is positive, and decreases as 
the scale decreases.
The long--dashed line corresponds to $n=1.2$, evaluated at
$N=N\sub{COBE}$.
The allowed region lies below this line and above the 
dotted line $\sigma = - c$.
For parameters lying above the short--dashed line (P--LB) 
the spectral index scale dependence will be detected by 
the Planck satellite.}
\end{figure}

\begin{figure}
\centering
\leavevmode\epsfysize=6.5cm \epsfbox{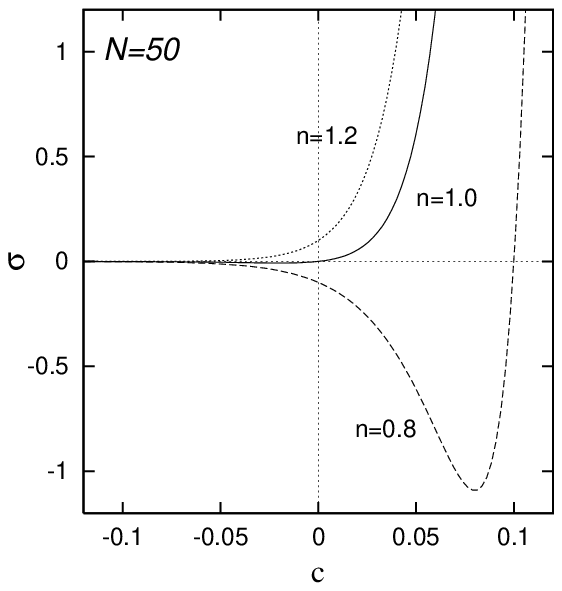}
\epsfysize=6.5cm \epsfbox{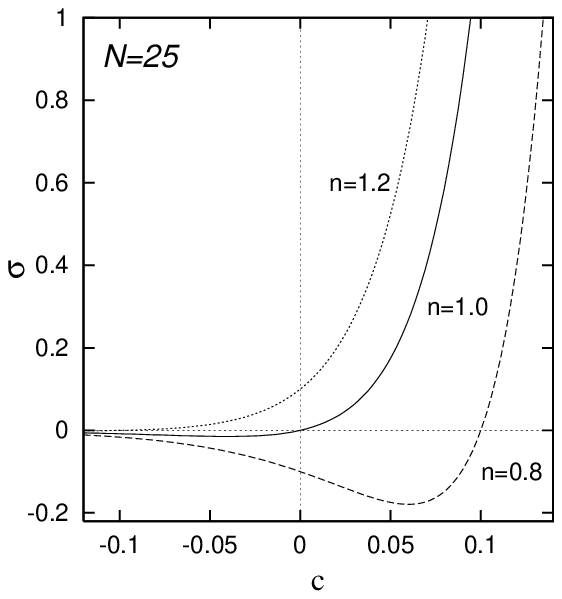}\\
\caption[allmodels]{\label{f:allmodels}
The lines $n=0.8$, $n=1.0$ and $n=1.2$ are shown, for $N=50$
and $N=25$. Each quadrant corresponds to one of the four models
of inflation. In the regime $c<0$, the line $n=1.0$ falls marginally 
below the line $\sigma=0$.}
\end{figure}

The range of observable scales corresponds to
\be
N\sub{COBE} - 10 \lsim N \lsim N\sub{COBE} \,.
\ee
In the case that $n$ is roughly constant over this range,
it is constrained by observation to $|n-1|/2 <0.1$.
As a rough estimate
we have applied this constraint at the two values
$N\sub{COBE}$ and $N=N\sub{COBE} -10$. The results are shown in
Figures \ref{f:modelone} to \ref{f:modelfour}.
An alternative view is provided by Figure \ref{f:allmodels},
in which the lines $n-1=0.1$, $0.0$ and $-0.1$ are shown,
for different values of $N$. Each quadrant corresponds to one
of the models; top right is model (i), bottom right is model (ii),
bottom left is model (iii) and top left is model (iv).

We have yet to consider the 
COBE normalization \eq{thiscobenorm}.
According to the above bounds, it 
is roughly
\bea
\frac{V_0^{1/2}}{\mpl^2}& \sim &
(10^{-4}{\rm \ to \ }10^{-5} ) e^{-\tau/|c|} \\
&\equiv& (10^{-4}{\rm \ to \ }10^{-5} ) (\phi_*/\mpl) \,.
\label{roughcobe}
\eea

Imposing only the requirement that inflation ends before 
nucleosnynthesis, one requires $V_0^{1/4} > 10\MeV$
corresponding to\footnote
{The upper limit corresponds to the condition $\mpl|V'/V|\ll 1$
for slow-roll inflation, which implies $V_0^{1/4}\lsim 10^{16}\GeV$.}
\be
0.012 \lsim |c| /\tau \lsim  1 \,.
\ee
But if the supersymmetry breaking scale during inflation is the same as 
in the true vacuum one will have
\be
10^5\GeV \lsim V_0^{1/4} \lsim 10^{10}\GeV \,.
\ee
The upper limit corresponds to gravity-mediated susy
breaking in the true vacuum, while 
the rest of the range corresponding to gauge-mediated
susy breaking in the true vacuum.\footnote
{In the latter case, the inflaton
sector can have only gravitational strength couplings 
with the visible sector.}
This corresponds to
\be
 0.019 \lsim |c|/\tau \lsim 0.033\,.
\ee

Finally, we should mention that in some cases,
a strong constraint is imposed by the requirement that excessive black 
hole formation does not occur soon after inflation ends.
With some assumptions, the
requirement for this 
\cite{cllw} is 
that the {\em rms} perturbation smoothed over a Hubble scale does not 
exceed $.04$ at the end of inflation. With further assumptions, this
is equivalent \cite{gl}
to  
\be
\delta_H < 0.01 \\,
\label{bhconstraint}
\ee
at the end of slow-roll inflation.

As the linear approximation ceases to be accurate before the end of 
inflation, the black hole constraint cannot in general be evaluated 
in terms of the
parameters $c$, $\sigma$ and $\tau$.
We examined the constraint in a case where it is likely to be 
particularly strong, namely the case of model (i) with inflation ending
at $\phi\sub{fast}$. We considered the case where only a single,
asymptotically free gauge group is relevant, using the formulas given 
in the next section. 

What we found was that $\delta_H$ increases very
quickly as $\phi\sub{fast}$ is approached.
If \eq{bhconstraint} is 
imposed precisely at the point
$\phi\sub{fast}$ 
defined by $\eta(\phi\sub{fast})=1$, it is a strong constraint.
But if we make the equally reasonable estimate
$\eta(\phi\sub{fast})=\frac12$ for the point where 
slow-roll inflation ends, and impose \eq{bhconstraint}
there instead, it  ceases to be a significant 
constraint. The conclusion, at least in this case,
is that a more careful calculation 
of black hole formation is required, than any appearing in the
literature.

\section{The case of a single gauge coupling}

\label{s4}

The Renormalization Group Equations can easily be solved 
in the case where only a single gauge coupling is relevant.
The result is \cite{ewanloop2,p98laura}
\be
m^2(\phi) = m^2_0 + {2 c\over b}\tilde m^2_0
{ \left[ 1- {1 \over \left[ 1-{b\alpha_0\over 2\pi}
\ln(\phi/\mpl) \right]^2}\right] } \,.
\ee
Here $m_0$ is the inflaton mass,
$\tilde m_0$ is the gaugino mass, $\alpha_0$ is the gauge 
coupling, all evaluated at the Planck scale, and $b$ and 
$c>0$ are numerical factors depending on the 
gauge group and its representations.

 We want the magnitude of $m^2$ to 
decrease as one goes down from the Planck scale.
This requires $m^2_0<0$, corresponding to model (i) or model (ii). 
Only model (i) in the case of asymptotic freedom ($b<0$) has been 
studied so far \cite{ewanloop2,p98laura}.

We evaluate $c$, $\sigma$ and $\tau$ to leading order in $\alpha$,
which is presumably all that is justified in a one-loop calculation.
It is convenient to use the following definitions, taken from
\cite{p98laura} but now applied to models (i) and (ii).
\bea
\mu^2 &\equiv& -m^2\mpl^2/V_0 \,,\\
A &\equiv & - {2 c\over b} {\tilde m^2\mpl^2\over V_0}
\label{param} \,,\\
\tilde \alpha &\equiv& {-b\alpha \over 2\pi } \,,\\
y&\equiv & \[1+\tilde \alpha_0 \ln(\phi/\mpl)\]^{-1} \,,\\
y_{**} &\equiv& \sqrt{1+{\mu^2_0\over A_0}} \,, \label{ystar}
\eea
where the subscript $0$ for a running quantity denotes the value
of that quantity at the Planck scale.
Applying the linear approximation one finds
\bea
c &=& 2 y_{**}^3 A_0 \tilde \alpha_0 
\label{i-c}\\
\tau &= & 2 A_0 y_{**}^2 (y_{**} -1 ) \,.
\label{i-tau}
\eea

We shall investigate models (i) and (ii) for arbitrary $b$, and in both 
cases assume that inflation continues until $|\mu^2|=1$.

\subsection{Model (i)}

Assuming that the mass continues to run until the end of inflation,
the condition 
\linebreak
$N(\phi\sub{end})=0$ gives
\be
\ln \sigma =  2 y_{**}^2 \( y_{**}^{-1} - y\sub{end}^{-1} \)
+\ln \[ \frac{4A_0 y_{**}^2 \( y\sub{end} - y_{**} \) }
{y\sub{end} + y_{**} } \] \,,
\label{i-sigma}
\ee
with
\be
y\sub{end} \equiv \( y_{**}^2 + A^{-1}_0 \)^{1/2}\,.
\label{yend}
\ee

This expressions are valid both in the case of negative $b$ (i.e. positive
$A_0$ and $\tilde \alpha_0$ ) and of positive $b$ (i.e. negative
$A_0$ and $\tilde \alpha_0$ ).

\begin{figure}
\centering
\leavevmode\epsfysize=5.3cm \epsfbox{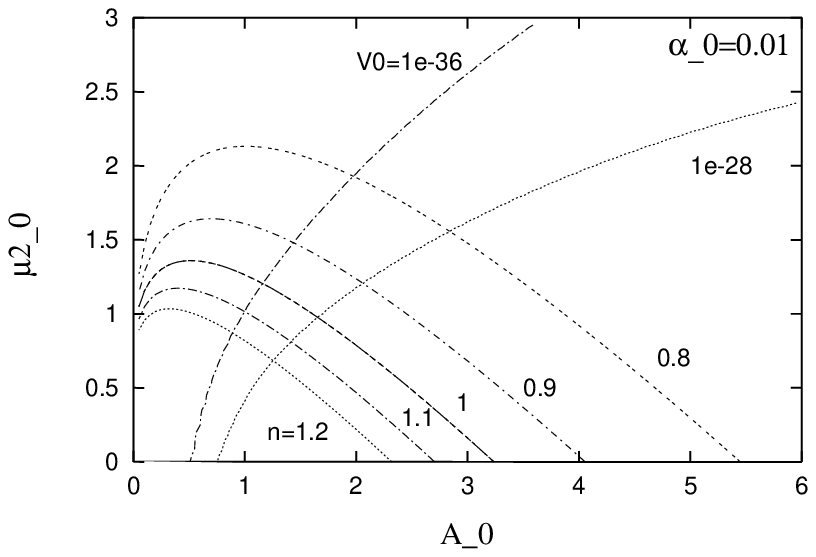}
\epsfysize=5.3cm \epsfbox{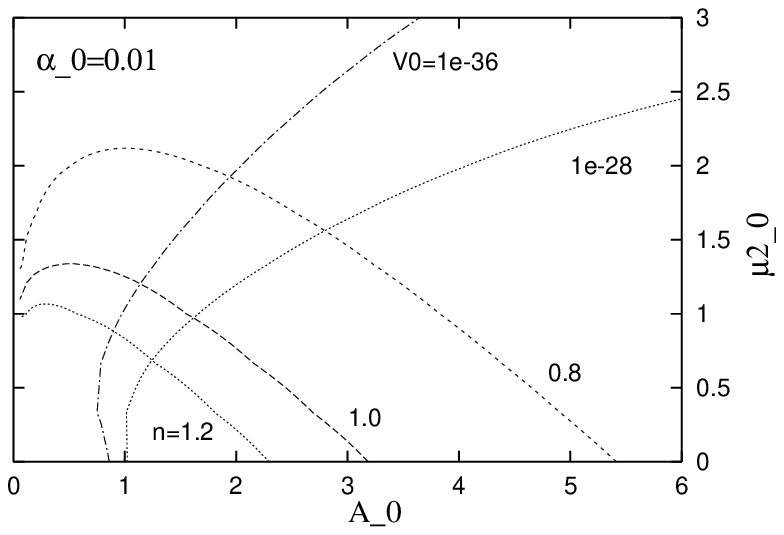}\\
\caption[comparison]{\label{f:comparison}
Lines of constant $n$ and of constant $V_0/\mpl^4$, 
in the plane $A_0$--$\mu_0^2$, for
$N\sub{COBE}=45$ and $\tilde \alpha_0 = 0.01$. The right figure
has been obtained using the linear approximation and equations 
(\ref{i-c}),(\ref{i-tau}) and (\ref{i-sigma}), while the left figure 
shows the exact result given in \cite{p98laura}.
For this value of $N\sub{COBE}$, 
\eq{bound-N-V0}
requires $V_0/\mpl^4\gsim 10^{-36}$, so the regime above that line
is forbidden.}
\end{figure}

In Figure \ref{f:comparison} we make a comparison between the 
present formalism, and the exact results of \cite{p98laura} for
the case of asymptotic freedom.
Lines of constant $n$ and of constant $V_0$ are shown, and the agreement 
is seen to be very good. In order to make a direct comparison with 
\cite{p98laura}, the lines of constant $n$ have all been evaluated
at $N=N\sub{COBE}$. The allowed region, as defined in the previous
Section, lies between the line $n=0.8$ evaluated at $N=N\sub{COBE}$
and the line $n=1.2$ evaluated at $N=N\sub{COBE}-10$; the latter
actually lies a little below the line $n=1.0$ shown in the Figure.

\begin{figure}
\centering
\leavevmode\epsfysize=5.3cm \epsfbox{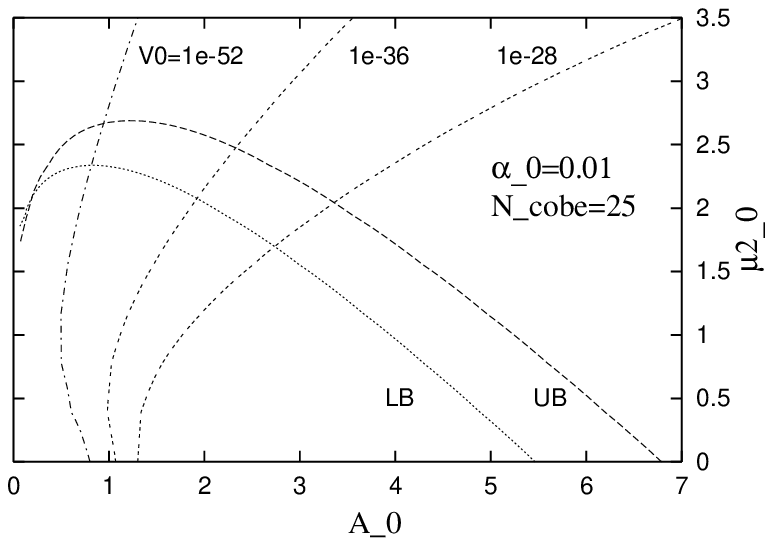}
\epsfysize=5.3cm \epsfbox{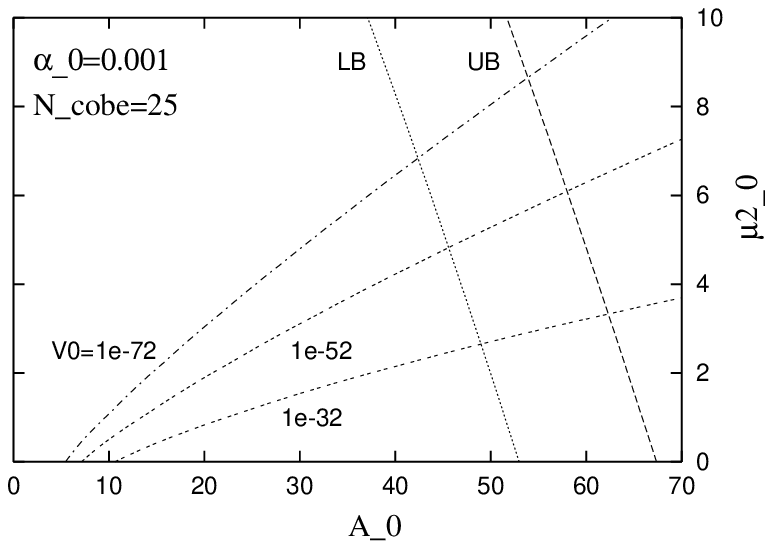}\\
\caption[model1-muA]{\label{f:model1-muA}
The contour lines in the linear approximation for the spectral index 
($n(N\sub{COBE})= 0.8$ and $n(N\sub{COBE}-10)=1.2)$ and $V_0$ for two different 
values of $\tilde\alpha_0 = 0.01, 0.001$ in the $\mu^2_0 - A_0$ plane 
for $N\sub{COBE}=25$. For this value of $N\sub{COBE}$, \eq{bound-N-V0}
requires $V_0/\mpl^4\gsim 10^{-72}$, so the regime above that line
is forbidden.}
\end{figure}

In Figure \ref{f:model1-muA} we give a similar plot
for the case $N\sub{COBE} = 25$, but this time plotting the line
$n=1.2$ at $N=N\sub{COBE} -10$ (curve marked LB). The allowed region is 
between this one and the line $n=0.8$ evaluated at $N=N\sub{COBE}$
(marked UB).
We 
can see that the general behaviour is similar to Figure \ref{f:comparison}.

Let us now consider the other case $b>0$, i.e. $A_0,\tilde\alpha_0 <0$ . 
As we can see from eq. (\ref{ystar}), the parameter space is limited to 
the region where $ |A_0| > \mu^2_0 $; this is due to the fact that
the running is weaker in this case and a large gaugino mass is necessary
to change the sign of the inflaton mass. The limiting case where
$ |A_0| = \mu^2_0 $ correspond to a vanishing mass in the asymptotic
limit $\phi = 0$, but such result is not reliable since the running would
surely stop before that point. 
Since we are actually setting $y\sub{end}$ by eq. (\ref{yend}), our parameter
space will be even more restricted by the condition 
$ |A_0| > \mu^2_0+1 $.

\begin{figure}
\centering
\leavevmode\epsfysize=5.3cm \epsfbox{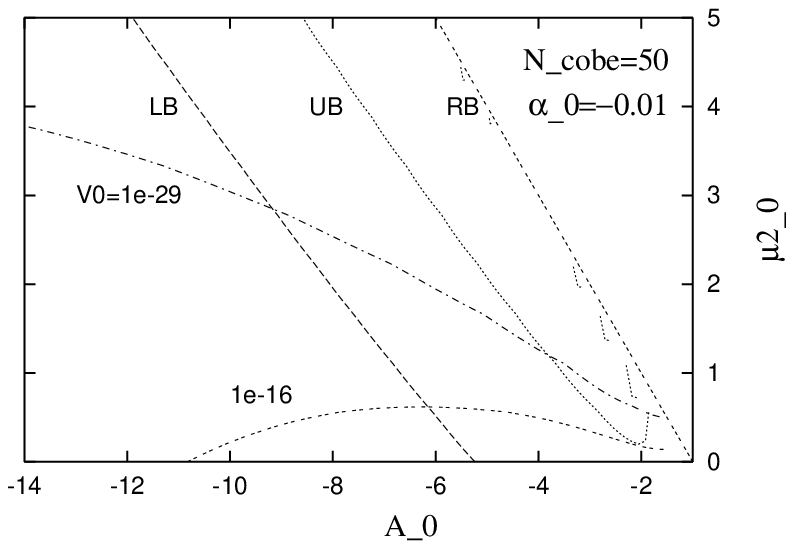}
\epsfysize=5.3cm \epsfbox{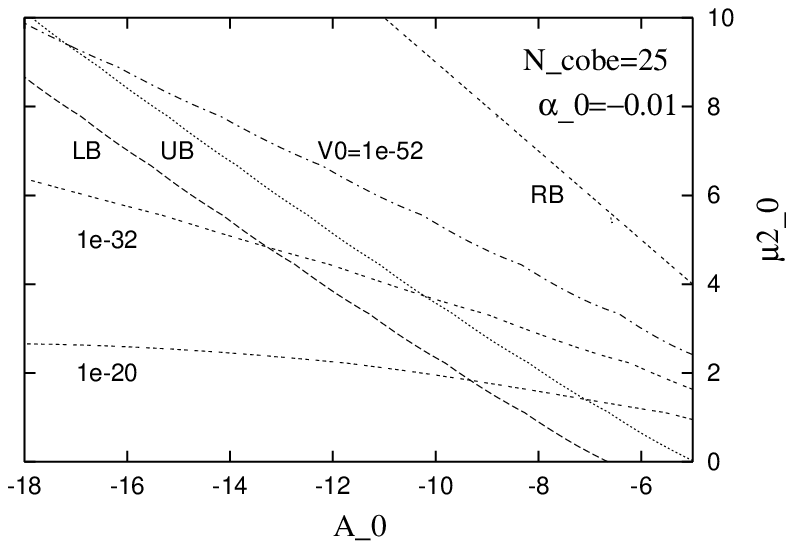}\\
\caption[model1neg]{\label{f:model1neg}
The contour lines in the linear approximation for the spectral index 
($n(N\sub{COBE})= 0.8$ and $n(N\sub{COBE}-10)=1.2)$ and $V_0$ for 
$\tilde\alpha_0 =-0.01$ in the $\mu^2_0 - A_0$ plane for $N\sub{COBE}=50, 25$. 
The line marked RB is the upper bound to the parameter space given
by requiring  $ |A_0| > \mu^2_0 +1$.
From \eq{bound-N-V0}, the regime
$V_0/\mpl^4\lsim 10^{-29}$ is forbidden if
$N\sub{COBE}=50$, and the regime
$V_0/\mpl^4\lsim 10^{-72}$ is forbidden if $N\sub{COBE}=25$.}
\end{figure}

In Figure \ref{f:model1neg} we show the allowed region for negative $A_0$
both for  $N\sub{COBE} = 50$ and $N\sub{COBE} = 25$ and for 
$\tilde\alpha_0 =-0.01$. It is bounded by the two lines UB, corresponding
to $n=1.2$ at $N=N\sub{COBE} -10$, and LB, corresponding to $n=0.8$
at $N=N\sub{COBE}$. Notice in this case a long narrow strip satisfies
the experimental constraints, but the larger values of $\mu^2_0$
are excluded because of the low value of $V_0$.

\subsection{Model (ii)}

Finally, we consider model (ii) in the same setting, i. e. considering that
the inflaton rolls on the right of the maximum ($\phi > \phi_*$) and
inflation ends when $|\eta| \simeq 1$. In this case $\sigma < 0$ and
is given by:
\be
\ln |\sigma | =  2 y_{**}^2 \( y_{**}^{-1} - y\sub{end}^{-1} \)
+\ln \[ \frac{4A_0 y_{**}^2 \( y_{**} - y\sub{end} \) }
{y\sub{end} + y_{**} } \] \,,
\ee
with
\be
y\sub{end} \equiv \( y_{**}^2 - A^{-1}_0 \)^{1/2}\,.
\ee

We see from the expression of $y\sub{end}$ that our parameter space
will in this case be restricted to the region where $A_0 +\mu^2_0 \geq 1$;
otherwise $|\eta| \simeq 1$ will never be reached and some other
mechanism should be responsible for the end of inflation.
Since we naturally expect $\mu^2_0$ to be of order 1, this condition
is not very restrictive, apart for negative $A_0$.

\begin{figure}
\centering
\leavevmode\epsfysize=5.3cm \epsfbox{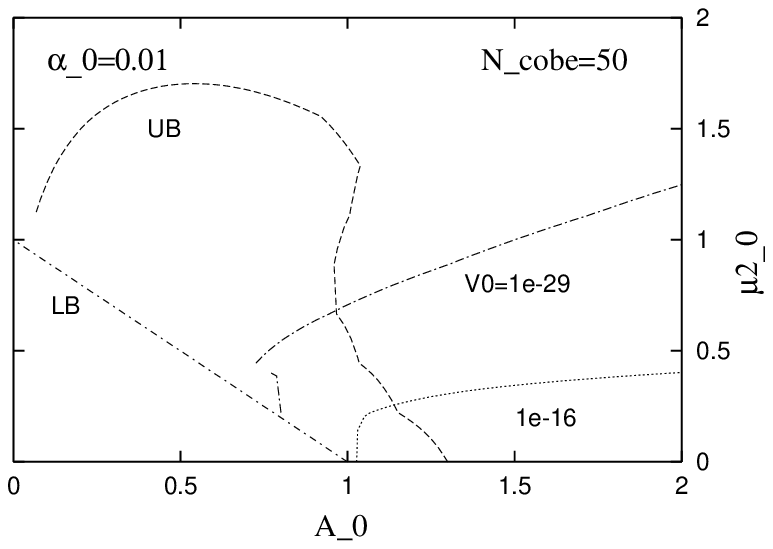}
\epsfysize=5.3cm \epsfbox{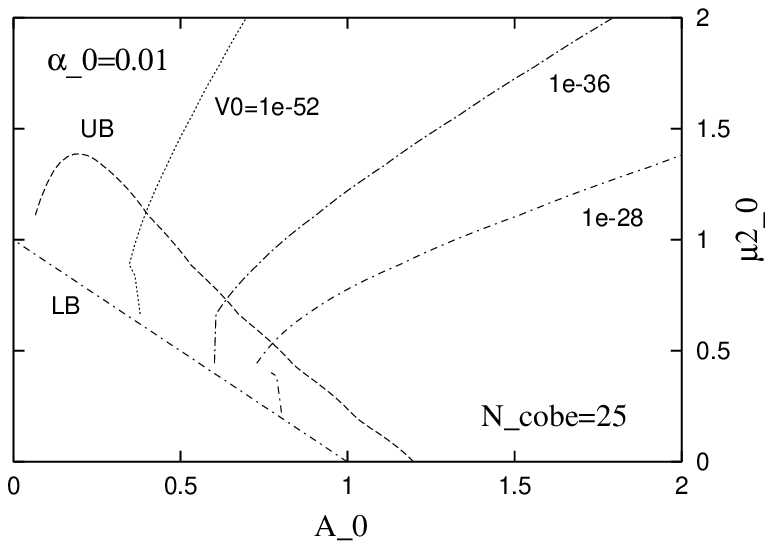}\\
\caption[model2-muA]{\label{f:model2-muA}
Contour lines for the spectral index 
at $N=N\sub{COBE} -10$ (line UB) 
and $V_0$ at $N\sub{COBE}$ for two different values of 
$N\sub{COBE}=50,25$ in the $\mu^2_0 - A_0$ plane for $\tilde \alpha_0 = 0.01$. 
This is a model of type (ii) and therefore $n-1<0$ in all the parameter
space; the experimentally allowed region is enclosed between the 
line UB and the line $A_0 +\mu^2_0 =1$ (LB).
From \eq{bound-N-V0}, the regime
$V_0/\mpl^4\lsim 10^{-29}$ is forbidden if
$N\sub{COBE}=50$, and the regime
$V_0/\mpl^4\lsim 10^{-72}$ is forbidden if $N\sub{COBE}=25$.}
\end{figure}

We see from Figure \ref{f:model2-muA} that in this model too,
there is a region of parameter space where $\mu^2_0$ and $A_0$
can have their expected values of order 1, with the reasonable
potential $10^5\GeV\lsim V_0^{1/4}\lsim 10^{10}\GeV$
($10^{-72}\lsim V_0/\mpl^4\lsim 10^{-32}$).\footnote
{This reverses the more pessimistic conclusion of Stewart
\cite{ewanloop1},
which was made on the basis of rough estimates only.}

We considered also the case of negative $A_0$ and $\tilde\alpha_0$,
but no region of the parameter space gives in this case an acceptable
spectral index, a part the line $|A_0| =\mu^2_0$, corresponding to
$\phi_* = 0$. Since, as we already discussed, we cannot extrapolate 
the running mass up to such low values of the field, this kind of
model is ruled out for the non asymptotically free case. 
Notice moreover that the large negative $n$ is due to the fact that
almost all the parameter space correspond to large $|\sigma|$ 
because of the positive exponential $\exp(1-y_{**}/y\sub{end})$ 
(negative for $b<0$). Therefore it seems improbable that considering
another mechanism for the end of inflation could change this result.

\section{Conclusion}

\label{s5}

A model of inflation with a running inflaton mass may contain many 
parameters, corresponding to the masses and couplings of the particles
responsible for the loop correction to the inflaton potential.
We have seen how to calculate the observable predictions in terms
of just three parameters, which in a given model can be calculated in 
terms of the masses and couplings. 

Observation presently allows a wide range of parameter space, which we 
have partially delineated. In the fairly near 
future, the allowed region will become much
smaller, because these models 
typically predict a spectral index $n$ with the distinctive
form $\frac12(n(k)-1) =\sigma e^{-cN} -c$. In this formula, 
the variation in wavenumber $k$ is given by $dk=-dN$,
and the parameters typically satisfy $|c|\lsim |\sigma|\lsim 1$.
If this model is correct, the scale dependence of $n(k)$ will
be detected by Planck in a large region of the parameter space.

On the theoretical side, the exploration of running-mass models
is only just beginning, and many questions remain unanswered regarding 
their possible relation with other aspects of physics beyond the 
Standard Model. In particular, one would like to know if the inflaton
can belong to the visible sector, or whether it must belong to a hidden
sector consisting of particles that have only gravitational-strength
coupling with particles possessing the gauge interactions of the 
Standard Model.

\section*{Acknowledgements}
We acknowledge useful conversations with Andrew Liddle, Antonio Riotto,
Graham Ross and Subir Sarkar. LC is supported by PPARC grant
GR/L40649.

\def\NPB#1#2#3{Nucl. Phys. {\bf B#1}, #3 (19#2)}
\def\PLB#1#2#3{Phys. Lett. {\bf B#1}, #3 (19#2) }
\def\PLBold#1#2#3{Phys. Lett. {\bf#1B} (19#2) #3}
\def\PRD#1#2#3{Phys. Rev. {\bf D#1}, #3 (19#2) }
\def\PRL#1#2#3{Phys. Rev. Lett. {\bf#1} (19#2) #3}
\def\PRT#1#2#3{Phys. Rep. {\bf#1} (19#2) #3}
\def\ARAA#1#2#3{Ann. Rev. Astron. Astrophys. {\bf#1} (19#2) #3}
\def\ARNP#1#2#3{Ann. Rev. Nucl. Part. Sci. {\bf#1} (19#2) #3}
\def\mpl#1#2#3{Mod. Phys. Lett. {\bf #1} (19#2) #3}
\def\ZPC#1#2#3{Zeit. f\"ur Physik {\bf C#1} (19#2) #3}
\def\APJ#1#2#3{Ap. J. {\bf #1} (19#2) #3}
\def\AP#1#2#3{{Ann. Phys. } {\bf #1} (19#2) #3}
\def\RMP#1#2#3{{Rev. Mod. Phys. } {\bf #1} (19#2) #3}
\def\CMP#1#2#3{{Comm. Math. Phys. } {\bf #1} (19#2) #3}

\end{document}